\documentclass[11pt]{article}
\usepackage{amsmath}
\usepackage{amssymb}
\usepackage{amsfonts}
\usepackage{eucal}
\usepackage{epsfig}
\newcommand{\be}{\begin{equation}}
\newcommand{\bea}{\begin{eqnarray}}
\newcommand{\eea}{\end{eqnarray}}
\newcommand{\ba}{\begin{array}}
\newcommand{\ea}{\end{array}}
\newcommand{\ee}{\end{equation}}

\expandafter\ifx\csname mathbbm\endcsname\relax

\else

\fi

\renewcommand{\baselinestretch}{1}

\begin{document}

\title{An analytical formulation for soliton-potential dynamics}  \author{Kurosh Javidan
\\Department of physics, Ferdowsi university of Mashhad\\ 91775-1436 Mashhad Iran
\\E-mail:javidan@um.ac.ir
\\}
\setlength{\parindent}{0mm} 
\renewcommand{\baselinestretch}{1} 
\newcommand{\ph}{\vec{\phi}} \newcommand{\pha}{\phi_{a}}
\newcommand{\dmu}{\partial_{\mu}} \newcommand{\umu}{\partial^{\mu}}
\newcommand{\dnu}{\partial_{\nu}} \newcommand{\unu}{\partial^{\nu}}
\newcommand{\di}{\partial_{i}} \newcommand{\ui}{\partial^{i}}
\renewcommand{\dj}{\partial_{j}} \newcommand{\uj}{\partial^{j}} \hoffset =
-2cm \textwidth = 170mm
\maketitle
\abstract

An analytical model for the soliton-potential interaction is presented, by constructing a collective coordinate for the system. Most of the characters of the interaction are derived analytically while they are calculated by other models numerically. We find that the behaviour of the soliton is like a point particle 'living' under the influence of a complicated potential, that is a function of soliton velocity and the potential parameters. The analytic model does not have a clear prediction for the islands of initial velocities in which the soliton may reflect back or escape over the potential well.

\section{Introduction}
Topological solitons are widely use as models for the description of particles generated as nontrivial solutions of nonlinear field theories. Skyrmions are solitons which are used as a model of hadrons. Some solitonic solutions appear in two-dimensional Quantum chromodynamics ($QCD_{2}$). In bosonized $QCD_{2}$ these type of solutions emerge as describing baryons and quark solitons, respectively. The generalized sine-Gordon model arises as the low-energy effective action of bosonized $QCD_{2}$ for unequal quark mass parameters. Also in the strong-coupling limit the static classical soliton which describes a baryon in $QCD_{2}$ turns out to be ordinary sine-Gordon kink. Modeling of optical self focusing phenomena, magnetic fluxes in real Josephson junctions are examples from other branches of physics.

In the meantime, dynamical evolution of a soliton during the interaction with potentials is an important phenomenon from the mathematical point of view and also because of its applications. Most of the researches are in base of numerical studies because such systems are generally non-integrable. So it is clear that we need suitable models with analytic solutions to test the validity of such phenomenon and predict their behaviour.

In this paper an analytic model for the interaction of sine-Gordon solitons with defects is presented and the results are compared with numerical simulation outcomes from other models. So we need a brief review of the available models which presents in section 2. The analytic model is introduced and will be solved in section 3. Presented model will be compared with other models in section 4. The results for the soliton-barrier system are presented in section5. The results will be compared with the predictions of other models in this section too. In section 6 soliton-well system is discussed. Some conclusion and remarks will be presented in section 7.
\section{Models for soliton-potential systems}
\textbf{model 1}: The potential generally arises from medium properties. The effects of medium disorders and impurities can be added to the equation of motion as perturbative terms. In this method, scattering of a soliton by a single impurity has been modeled as\cite{r1,r2}
\begin{equation}\label{eqmotion}
\phi_{tt}-\phi{xx}+\left(1+\sigma\delta(x)\right)\frac{\partial U}{\partial \phi}=0
\end{equation}
where$\sigma$ denotes the strength of impurity and $\frac{\partial U}{\partial \phi}=sin\phi$ for the sine-Gordon model. For an attractive potential well,$\sigma$ is negative ($\sigma <0$ ) and for a barrier $\sigma$ is a positive number ($\sigma >0$).

The impurity has been added as an external potential in this model . The interaction can be analyzed in term of some degree of freedom for the soliton ( position of the center of the soliton) and an impurity mode for the external potential.

In this approach the impurity causes the interaction of a soliton with an effective potential. In particular soliton can be trapped by an attractive potential because of energy loss due to radiation. In this model the impurity is not a rigid object. It has a localized oscillating state,so-called impurity mode.

In the absence of the impurity ($\sigma =0$ ) equation (1) has an exact one soliton solution as
\begin{equation}\label{one soliton}
\phi_{k}=4 \arctan \left ( \exp \left ( \frac {x-X(t)}{\sqrt {1-V^2}} \right ) \right )
\end{equation}
where $X(t)=X_{0}-Vt$ and V is the soliton velocity.

If we linearize equation around its ground state, we have
\begin{equation}\label{linear}
\phi_{tt}-\phi{xx}+\left(1+\sigma\delta(x)\right)\phi=0
\end{equation}
which has a localized oscillating mode
\begin{equation}\label{impurity}
\phi_{impurity}(x,t)=a(t)\exp \left (-\sigma \frac{\left|x\right|}{2} \right )=0
\end{equation}

Two dynamical variables, X(t) and a(t) explain the dynamics of the soliton-potential system. One can describe the soliton-impurity interaction by substituting $\phi=\phi_{k}+\phi_{impurity}$ into the lagrangian of the system and integrating over the variable 'x' [2]. After that, the kink coordinate X(t) and impurity mode a(t) are considered as collective coordinate variables and their evolution describe the situation of the soliton during the interaction. Therefore the soliton is changed to a point particle with an effective mass of $ m_{eff} =8$ in the effective potential $V(X)=\frac{2\sigma}{\cosh^2{X}}$. This potential creates the effective force
\begin{equation}\label{potential}
F(X)=-V^\prime(X)=\frac{4\sigma sinh(X)}{cosh^{3}(X)}
\end{equation}
 
 \textbf{Model 2}: The effects of the potential also can be taken into account by making some parameters of the equation of motion (or lagrangian) to be as functions of space or time \cite{r3,r4}. In this approach, a finite size, finite strength potential is included by appropriately modifying the coefficient of the nonlinear term in the lagrangian or equation of motion. The effective lagrangian from this model for the sine-Gordon soliton-potential system is
\begin{equation}\label{wojtek}
{\cal L}=\partial_ {\mu}\phi\partial^{\mu}\phi-\lambda ^{2}\left ( 1-cos\phi\right )
\end{equation}
with solution
\begin{equation}\label{one soliton wjz}
\phi_{k}= 4\arctan \left ( \exp \left (\lambda \frac {x-X(t)}{\sqrt {1-V^2}} \right ) \right )
\end{equation}
$\lambda$ is chosen as
\begin{equation}\label{lambda}
\lambda= \left\{
\begin{array}{clrr}
1 & \left|x\right|>p\\\lambda_{0} & \left|x\right|<p
\end{array}\right\}
\end{equation}
where p is the width of the potential. For $\lambda_{0}<1$ we have a potential well and $\lambda_{0}>1$ describes a potential barrier. A delta-like potential with the strength of $\epsilon_{0}$ is constructed with the constraint $\lambda_{0}p=\epsilon_{0}$.

\textbf{Model 3}: One can add such effects to the lagrangian of the system by introducing a suitable nontrivial metric for the back ground space-time, without missing the topological boundary conditions \cite {r5,r6,r7,r8}. In other words, the metric carries the information of the medium. The general form of the action in an arbitrary metric is:
\begin{equation}\label{action metric}
I=\int{{\cal L}(\phi , \partial_{\mu}\phi)\sqrt{-g}d^{n}x dt }
\end{equation}
where "g" is the determinant of the metric $g^{\mu \nu} (x)$. Energy density of the "field + potential" can be found by varying "both" the field and the metric \cite {r7}. For the lagrangian of the form
\begin{equation}\label{sg lagrangian}
{\cal L}=\frac{1}{2}\partial_{\mu}\phi\partial^{\mu}\phi-U(\phi)
\end{equation}
the equation of motion becomes \cite {r7,r9}
\begin{equation}\label{Eq motion}
\frac {1}{\sqrt{-g}}\left (\sqrt{-g}\partial_{\mu}\phi\partial^{\mu}\phi+\partial_{\mu}\phi\partial^{\mu}\sqrt{-g}\right )+\frac {\partial U(\phi)}{\partial \phi}=0
\end{equation}
The suitable metric in the presence of a weak potential V(x) is \cite {r5,r6,r7}:
\begin{equation}\label{metric}
g_{\mu \nu}(x)\cong\left(
\begin{array}{clrr} 1+V(x) & 0 \\ 0 & -1
\end{array}\right)
\end{equation}
The equation of motion (11) (describes by Lagrangian (10)) in the background space-time (12) is
\begin{equation}\label{Equation motion}
\left ( 1+V(x)\right )\frac {\partial^{2}\phi}{\partial t^2}-\frac {\partial^{2}\phi}{\partial x^2}-\frac {1}{2\left|1+V(x)\right|}\frac {\partial V(x)}{\partial x}\frac {\partial \phi}{\partial x}+\frac {\partial U(\phi)}{\partial \phi}=0
\end{equation}
For the sine-Gordon model, we have $U(\phi)=1-\cos \phi$. A potential of the form $V(x)=a e^{-b(x-c)^{2}}$  has been chosen in \cite {r7} while a square shape potential has been used for simulations in \cite {r8}. In the above potential, parameter "a" controls the strength of the potential, "b" represents its range, and "c" indicates the center of the potential. If $a>0$, the potential shows a barrier and for $a<0$ the potential acts as a potential well.

 \section {collective coordinate system for model 3}

The center of a soliton can be considered as a particle, if we look at this variable as a collective coordinate. The collective coordinate could be related to the potential by using one of the above models. The third model is able to give us analytic solution for the evolution of the soliton center during the soliton-potential interaction.

Here we work on the sine-Gordon model with its one soliton solution of (2). By inserting the solution (2) in the lagrangian (10) and using metric (12), with adiabatic approximation \cite {r1,r2} we have
\begin {equation} \label {lag}
{\cal L}=\frac {1}{2}\left( \sqrt{-g}\right)^{3} \frac {4\dot{X}^{2}}{\cosh^{2}\left(x-X(t)\right)}-\sqrt{-g}\frac {4}{\cosh ^{2}\left(x-X(t)\right)}
\end {equation}
For the weak potential V(x) (14) becomes
\begin {equation} \label {lag app}
{\cal L}\approx \left(1+\frac{3}{2}V(x)\right)\frac {2\dot{X}^2}{\cosh ^{2}\left(x-X(t)\right)}-\left(1+\frac{1}{2}V(x)\right) \frac {4}{\cosh ^{2}\left(x-X(t)\right)}
\end {equation}
X(t) remains as a collective coordinate if we integrate (15) over variable x
\begin {equation} \label {L}
L=\int{{\cal L}dx}=4\dot{X}^2 +3\dot{X}^2\int{\frac {V(x)dx}{\cosh ^{2}\left(x-X(t)\right)}}-8-2\int{\frac {V(x)dx}{\cosh ^{2}\left(x-X(t)\right)}}
\end {equation}
The equation of motion for the variable X(t) results from the (16)
\begin {equation} \label {X Eq Motion}
8\ddot{X}+6\ddot{X}\int{\frac {V(x)dx}{\cosh ^{2}\left(x-X(t)\right)}}+\left(6\dot{X}^2+4\right)\int{\frac {V(x)\sinh\left(x-X(t)\right)dx}{\cosh^3\left(x-X(t)\right)}}=0
\end {equation}
It is a general equation for the any kind of potential. If we take the potential $V(x)=-\epsilon \delta(x)$ then (17) becomes
\begin {equation} \label {X Eq Motion2}
8\ddot{X}\left(1-\frac{3\epsilon}{4\cosh^2X}\right)+\left(\frac{3\dot{X}^2}{2}+1\right) \frac{4\epsilon\sinh X}{\cosh^3 X} =0
\end {equation}
The above equation shows that the energy peak of the soliton moves under the influence of a complicated force which is function of its position and velocity. Note that an effective force in the form of equation (5) of model 1 appears in equation (18) when the soliton velocity is small $(\dot{X}\rightarrow 0)$. If $\epsilon>0$ we have a barrier and $\epsilon<0$ creates a potential well.
The energy of the soliton in the presence of the potential becomes
\begin {equation} \label {Energy}
E=4\dot{X}^2+\frac{3\epsilon\dot{X}^2}{\cosh^2X}+8+\frac{2\epsilon}{\cosh^2X}
\end {equation}
When the soliton is far from the center of the potential ($X\rightarrow\infty$) (19) reduces to $E=4\dot{X}^2+8$. It is the energy of a particle with a mass of 8. Some of the features of the soliton behaviour can be found from the (19). For example, suppose that a potential barrier of height $\epsilon$ is located at the origin. A soliton with a low velocity reflects back from the barrier and a high energy soliton climbs over the barrier and passes over it. So we have a critical value for the velocity of the soliton which separates these two situations. The energy of a soliton in the origin (X=0) comes from (19)$E(X=0)=\left(4+3\epsilon\right)\dot{X}^2+8+2\epsilon$. The minimum value of the energy for a soliton in this situation is $E=8+2\epsilon$. On the other hand, a soliton which comes from the infinity with initial velocity $v_{c}$ has the energy of $E\left(X=\infty\right)=4v_{c}^2+8$. It is clear that it can pass though the barrier if $v_{c}>\sqrt{\frac{\epsilon}{2}}$.

Equation (18) has an exact solution as follows
\begin {equation} \label {xxdot}
\frac{3\dot{X}^2+2}{3\dot{X_{0}}^2+2}=\frac{\cosh^2X \left(\cosh^2X_{0}+\frac{3\epsilon}{4} \right)}{\cosh^2X_{0} \left(\cosh^2X+\frac{3\epsilon}{4} \right)}
\end {equation}
where $X{0}$ and $\dot{X}_{0}$ are initial position and initial velocity respectively. Many of the characters of soliton-potential system can be extracted from the above solution. In the next sections some results are discussed and also compared with the results of the other models.

\section {Comparing of the models}

These three models can be compared numerically. All these models (for a delta-like potential) have a parameter in their equation of motion,$\sigma$ in model 1, $\lambda_{0}$ in model 2 and $\epsilon$ in model 3. The parameters control the strength of the external potential.

It is possible to compare these three parameters in a specific situation by simulation and adjusting parameters to have same results by different models for that specific situation. It is expected to find approximately the relation between the parameters in other situations. A set of simulations for the three models have been performed for finding $v_{c}$ with respect to different values of the potential strength with using three models. It is observed that model 3 predicts the value of $v_{c}=\sqrt{\frac{\epsilon}{2}}$ when the soliton is far from the center of the potential ($X_{0}\rightarrow\infty$). Simulations using models 1 and 2 show the same $\sqrt{\frac{\epsilon}{2}}$ behaviour. An effective strength is found by interpolation of simulation results on the $\sqrt{\frac{\alpha+\beta\epsilon}{2}}$ for both models 1 and 2 with respect to parameter of model 3. Figure 1a shows the results of simulations for (1) with sine-Gordon model. Effective strength of model 1 with respect to model 3 is
\begin {equation} \label {barazesh}
\epsilon_{1}=(0.0275\pm0.0022)+(0.786\pm0.0064)\sigma
\end {equation}
with standard deviation of $8.5\times10^{-6}$. Figure 1b presents the results of the fitting for the model 2. The result of the fitting is
\begin {equation} \label {fit}
\epsilon_{2}=(-0.0645\pm0.00221)+(0.8004\pm0.0065)\lambda_{0}
\end {equation}
with standard deviation of $6\times10^{-5}$

Figures 1 show that the three models are in agreement with each other if, $v_{c}=\sqrt{\frac{\epsilon_{effective}}{2}}$ where the effective parameters are calculated for models 1 and 2 with respect to parameter of the model 3.
Simulations have been done using Ronge-Kutta method for time derivatives and finite difference method for space derivatives. Space grids have been chosen $\Delta x=0.01,0.05$ and some times 0.025. Time cells have been chosen $\Delta t=\frac{\Delta x}{4}$ in the simulations. Delta function was simulated by the function $\sqrt{\frac{\alpha}{\pi}}e^{-\alpha x^2}$ with several values for $\alpha$.

\begin{figure}[htbp]
  \begin{center}
    \leavevmode
 \epsfxsize=17cm   \epsfbox{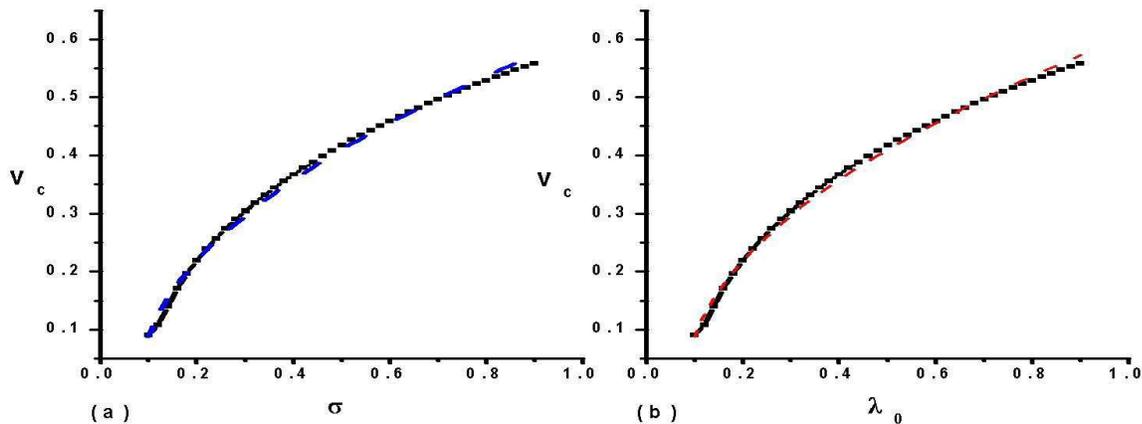}
  \end{center}
  \caption{The critical velocity $v_{c}$ respect to strength of the potential. Figure 1a presents the results of model 1 and figure 1b for model 2.Dotted plots denotes the fitted curve on the function $\sqrt{\frac{\alpha+\beta\epsilon}{2}}$ and solid lines with data points show the simulation results.}
  \label{fig:fig1}
\end{figure}

\section {Soliton-barrier system}

A soliton-barrier system is modeled with $\epsilon>0$ in (18) or (20). Consider a soliton with initial velocity of $\dot{X}_{0}$ at initial position of $X_{0}=-\infty$. Equation (20) shows that the soliton reaches the infinity again with the final velocity $\dot{X}=\pm\dot{X}_{0}$. The soliton goes to $-\infty (+\infty)$ if its initial velocity is less (more) than the critical velocity $v_{c}$. If the soliton is located at some position like $X_{0}$ (which is not necessary infinity) the critical velocity will not be $\sqrt{\frac{\epsilon}{2}}$. Neither model 1 nor model 2 has analytical prediction for the critical velocity in this situation. However we can investigate this situation numerically using these models. Now let us study the situation with model 3. The soliton can pass over the barrier if the soliton energy is greater than the energy of a static soliton at the top of the barrier. So a soliton in the initial position $X_{0}$ with initial velocity of $\dot{X}_{0}$ has the critical initial velocity if its velocity becomes zero at the top of the barrier $X=0$. Consider a soliton with initial conditions of $X_{0}$ and $\dot{X}_{0}$. If we set $X=0$ and $\dot{X}=0$ in equation (20) then $v_{c}=\dot{X}_{0}$. Therefore we have from (20)
\begin {equation} \label {vc}
v_{c}=\sqrt{\frac{\epsilon}{2}\frac{\cosh^2 X_{0}-1}{\cosh^2 X_{0}+\frac{3\epsilon}{4}}}
\end {equation}
Figure 2a presents the critical velocity as a function of initial position ($X_{0}$) with $\sigma =0.4, 0.6$ and 0.75 in model 1. The equivalent potential strength using model 3 from (21) is $\epsilon_{1}=0.3419, 0.4991$ and 0.6170 which are shown in the figure 2a with solid lines. Dotted lines in figure 2b show the critical velocity as a function of initial position ($X_{0}$) using model 2 with $\lambda_{0}=0.5, 1.0$ and 2.0. The solid lines in figure 2b show the equivalent situations using model 3 with $\epsilon_{2}=0.2357, 0.7359$ and 1.5363. These figures show that model 3 is in a very good agreement with model 1 and model 2.

\begin{figure}[htbp]
  \begin{center}
    \leavevmode
  \epsfxsize=17cm  \epsfbox{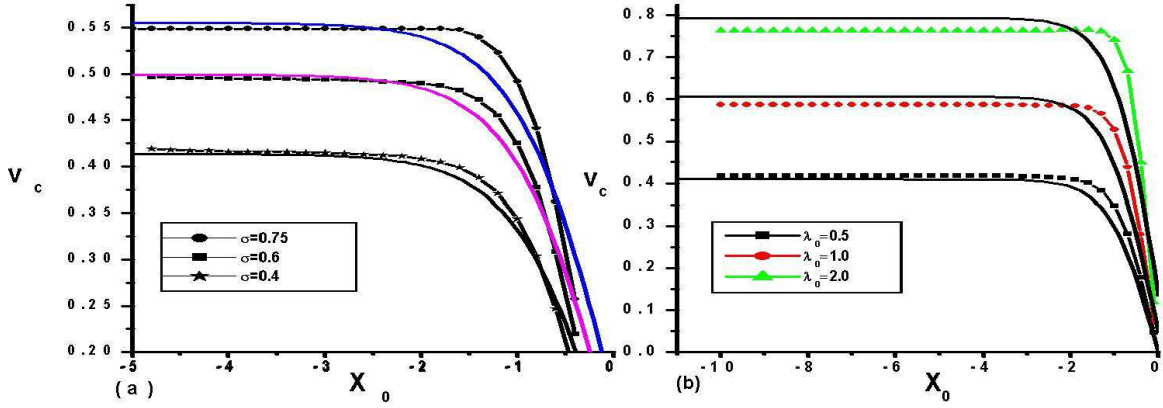}
  \end{center}
\caption {2a shows the Critical velocity $v_{c}$ respect to initial position $X_{0}$ simulated with model 1 and figure 2b presents the simulation results for model 2. Solid lines with data points present simulation results and dashed lines are plotted using equation (23) with equivalent effective strength calculated from (21) and (22). }
  \label{fig:fig2}
\end{figure}

If the initial velocity is less than $v_{c}$ then exists a return point in which the velocity of the soliton is zero. This point is derived from (20)
\begin {equation} \label {x stop infinity}
X_{stop}=\cosh^{-1}\left(\sqrt{\frac{3\epsilon}{2\alpha-4}} \right) ,  \alpha=\left( 3\dot{X}^2+2\right) \left(1+\frac{3\epsilon}{4\cosh ^2 X_{0}}  \right)
\end {equation}
where $X_{0}$ and $\dot{X}_{0}$ are initial position and initial velocity respectively. If the above equation is rearranged as
\begin {equation} \label{x stop}
\frac{1}{\cosh^2X_{stop}}=\frac{1}{\cosh^2X_{0}}+\dot{X}_{0}^2\left(\frac{2}{\epsilon}+\frac{3}{2\cosh^2 X_{0}} \right)
\end {equation}
one can find a linear relation between $\frac{1}{\cosh^2 X_{stop}}$ and $\dot{X}_{0}^2$. Figure 3a shows $\frac{1}{\cosh^2 X_{stop}}$ as a function of $\dot{X}_{0}^2$ with constant value for $X_{0}$ and some values of $\epsilon$ using model 1. All the simulations result the same value at $\dot{X}_{0}=0$ which is equal to $\frac{1}{\cosh^2 X_{0}}$. Equation (24) also shows another linear relation between $\frac{1}{\cosh^2 X_{stop}}$ and $\frac{1}{\cosh^2 X_{0}}$. Figure 3b demonstrates the numerical simulations with model 1 for this situation. Model 1 is in agreement with linear relation between $\frac{1}{\cosh^2 X_{stop}}$ and $\frac{1}{\cosh^2 X_{0}}$ as well as linear relation between $\frac{1}{\cosh^2 X_{stop}}$ and $\dot{X}_{0}^2$, which are conclude from the analytic model 3. Model 2 also show the same linear relations. 
\begin{figure}[htbp]
  \begin{center}
    \leavevmode
  \epsfxsize=18cm
   \epsfbox{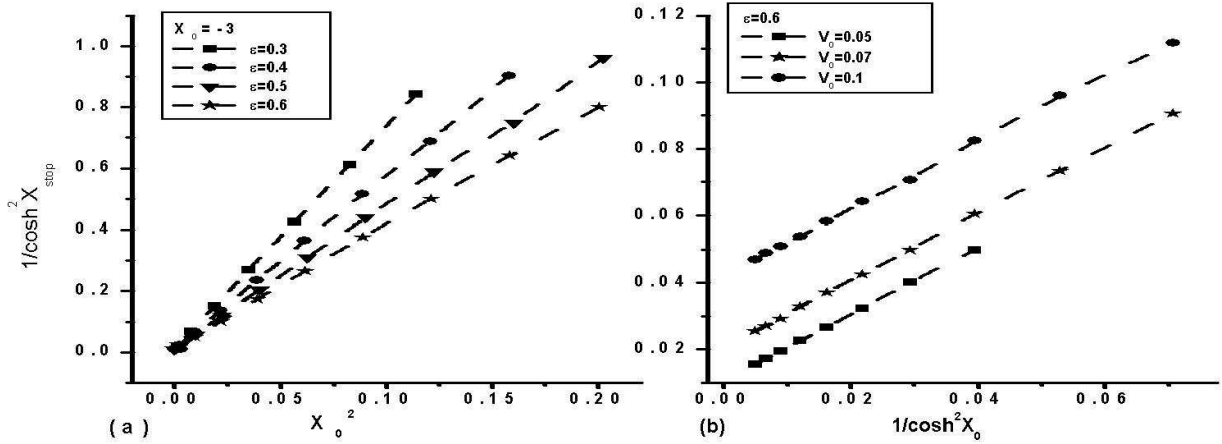}
  \end{center}
\caption {Figure 3a shows linear relation between $\frac{1}{\cosh^2 X_{stop}}$ and $\dot{X_{0}}^2$. Linear relation between $\frac{1}{\cosh^2 X_{stop}}$ and $\frac{1}{\cosh^2 X_{0}}$ has been shown in figure3b. Simulations have been done with model 1. }
  \label{fig:fig3}
 \end{figure}

The trajectory of a soliton during the interaction by the potential,X(t) follows from (20) as
\begin {equation} \label {t}
t=\int^{X(t)}_{X(t=0)}\left(\sqrt{ \frac{\left(3\dot{X}^2+2\right) \left( \cosh^2 X_{0}+\frac{3\epsilon}{4} \right)} {3\cosh^2 X_{0}} \frac {\cosh^2 X}{\cosh^2 X + \frac{3\epsilon}{4}  }-\frac{2}{3}  } \right)^{-1}dX
\end {equation}
The above integral has been evaluated numerically by using Rubmerg's method and X(t) was plotted versus t. This result was compared with direct simulation using model 1. Figure 4 shows the result for a system with $\epsilon =0.4$, $X_{0}=-5$ and $\dot{X}_{0}=-0.5$. There is a little difference between the predicted final velocities from different models after interaction. The difference is reduced when the height of the potential($\epsilon$) reduces. The difference is due to the approximation which is used for deriving (15) from (14).
\begin{figure}[htbp]
  \begin{center}
    \leavevmode
  \epsfxsize=10cm
    \epsfbox{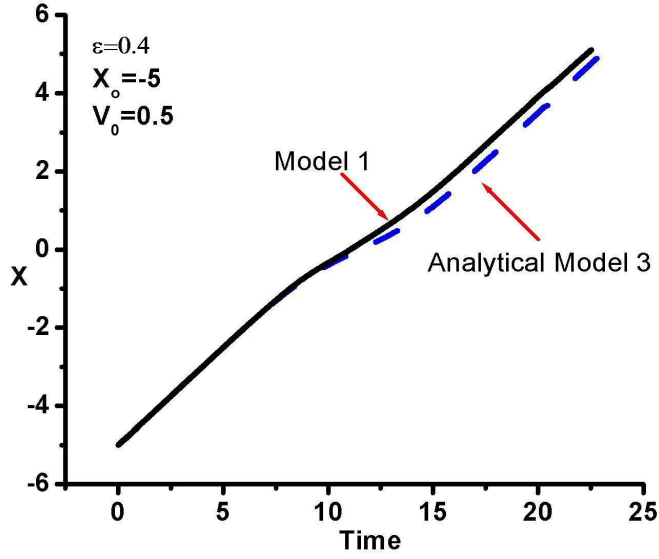}
  \end{center}
\caption {Trajectory of a soliton during the interaction with a delta-like potential. Solid line presents the result of model 1. Dashed lines show the same situation calculated with analytic model 3. }
  \label{fig:fig4}
\end{figure}
Same results have been found when the soliton reflects back after the interaction. Another interesting experiment is finding the time that a soliton needs to reach a fixed point when it has different initial velocities. This situation has been investigated with both model 1 and model 3. Figure 5 shows the results for some different soliton-potential systems.

\begin{figure}[htbp]
  \begin{center}
    \leavevmode
  \epsfxsize=10cm  \epsfbox{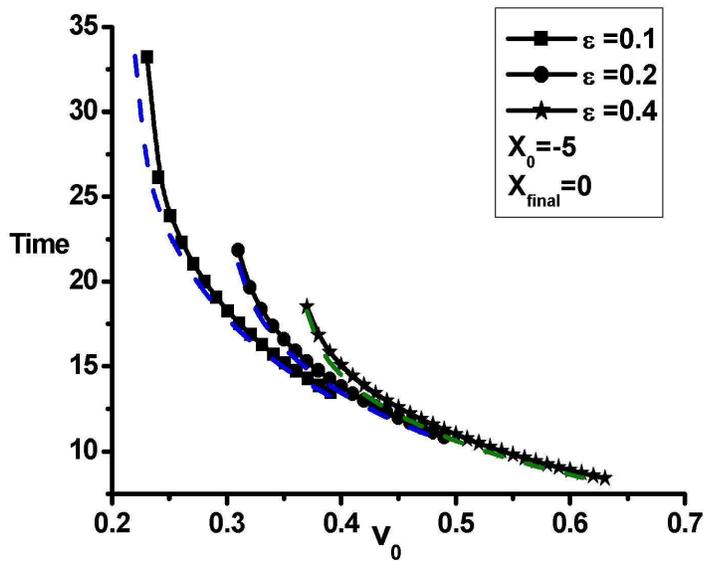}
  \end{center}
\caption {The time that a soliton at initial position $X_{0}=-5$ needs to reach the final position X=0 as a function of soliton initial velocity. Solid lines show the simulation results using model 1 and the dashed lines present same situation calculated with model 3.  }
  \label{fig:fig5}
\end{figure}
Several different simulations have been set up and the results of the three models compared. All the simulations show very good agreements between three models. Collation between model 2 and analytic model 3 also shows a good agreement between these two models, as well as what we can see between the model 1 and model 3. In order to avoid adding more figures, the results of model 1 has been reported. So we can conclude that the analytic model 3 can predict the characteristics of a soliton-barrier system.
It is concluded that the soliton 'lives' like a point particle but the extended nature of the soliton induces some effects on the potential and therefore the effective potential becomes more complicated than what we see in a point particle-barrier systems.

\section {Soliton-well system}

The soliton-well system is a very interesting problem. The behaviour of a soliton during the interaction with a potential well is very different from a point particle in the same situation. It is found that some differences can be explained by the characters of the effective potential.

Changing $\epsilon$ to $-\epsilon$ in (20) changes potential barrier to potential well. The solution for the system is
\begin {equation} \label {well eq}
\frac{3\dot{X}^2+2}{3\dot{X}_{0}^2+2}=\frac {\cosh^2X\left(\cosh^2X_{0}-\frac{3\epsilon}{4} \right)}{\cosh^2X_{0}\left(\cosh^2X-\frac{3\epsilon}{4} \right)}
\end {equation}
Let's examine the validity of (26) by simulating model 1. Models 1 and 2 have similar behaviour. Here the simulations are performed using only model 1.

Consider a potential well with the depth of $\epsilon$. A soliton at the initial position $X_{0}$ moves toward the well with the initial velocity of $\dot{X_{0}}$. It interacts with the potential and reaches a maximum distance from the center potential $X_{max}$. The velocity of the soliton at $X_{max}$ is zero. $X_{max}$ can be found from (24) but with $\epsilon<0$. Figure 6 shows the results of simulations with $\epsilon=-0.3,-0.2$ and -0.1. The dashed lines show the results of linear fitting on the simulation data with model 1 and the solid lines presents the results of model 3 with effective potentials from (21).

\begin{figure}[htbp]
  \begin{center}
    \leavevmode
  \epsfxsize=10cm  \epsfbox{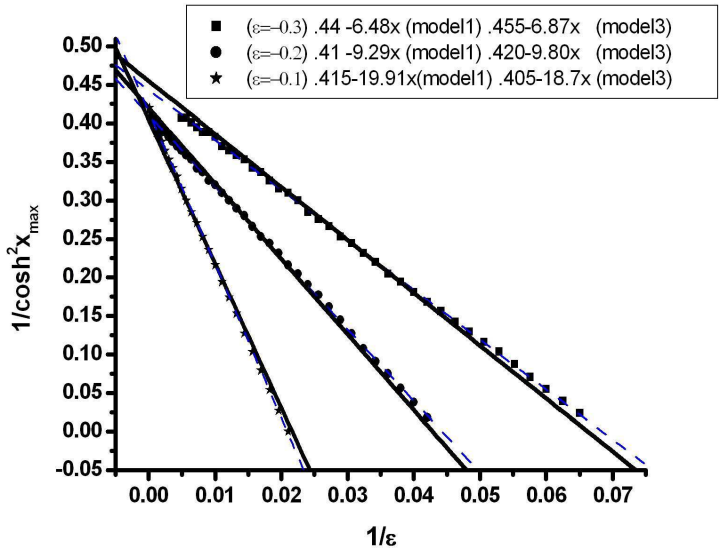}
  \end{center}
\caption { The dashed lines with data points show results of simulations using model 1 and the solid lines present same situation calculated with model 3. The effective potentials has been calculated with (21). }
  \label{fig:fig6}
\end{figure}
Simulations are in agreement with linear relation between $\frac{1}{\cosh^2X_{0}}$ and $\frac{1}{\epsilon}$. Also there is another linear relation between $\frac{1}{\cosh^2X_{0}}$ and $\frac{1}{\cosh^2X_{max}}$.

The time required for a soliton with an initial velocity $V_{0}$, from initial position $X_{0}$ to reach the origin also has been simulated using model 1 and has been calculated using model 3. Figures 7 show the results for a soliton with an initial position $X_{0}=-5$. The effective potential from (21) has been used in model 3.
\begin{figure}[htbp]
  \begin{center}
    \leavevmode
  \epsfxsize=17cm
  \epsfbox{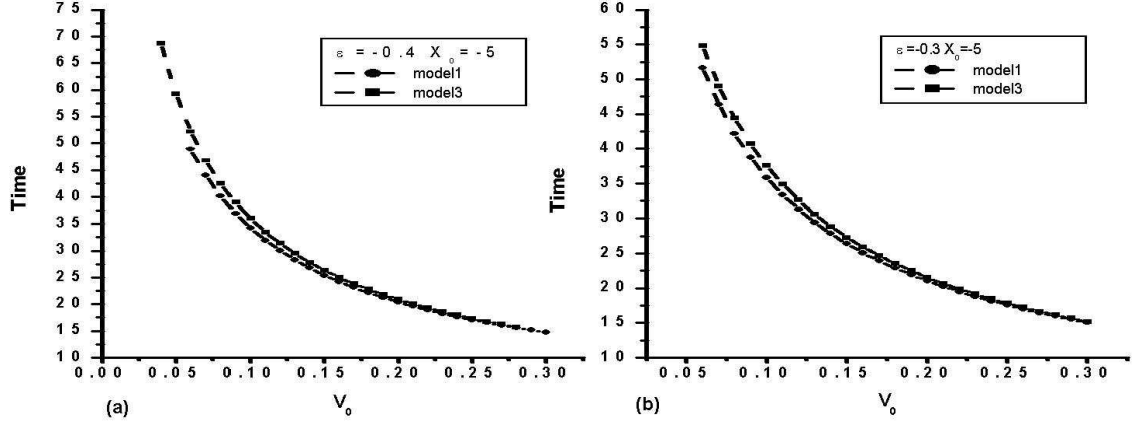}
  \end{center}
\caption {The time needed for a soliton to travel from initial position $X_{0}=-5$ to origin as a function of soliton initial velocity. Figure7a shows the simulation results using model 1 and same situation calculated with model 3. The strength of the potential is $\epsilon=-0.4$. Figure 7b presents the results for $\epsilon=-0.3$. }
  \label{fig:fig7}
\end{figure}

These results show that the equation (21) is valid for the potential well too. Also we see that the model 3 covers the bahaviour of soliton-well system with an acceptable precision.

A soliton can pass through the potential well if it has suitable initial velocity. Figure 8 presents the simulation results using model 1 and calculation using model 3.

\begin{figure}[htbp]
  \begin{center}
    \leavevmode
  \epsfxsize=10cm \epsfbox{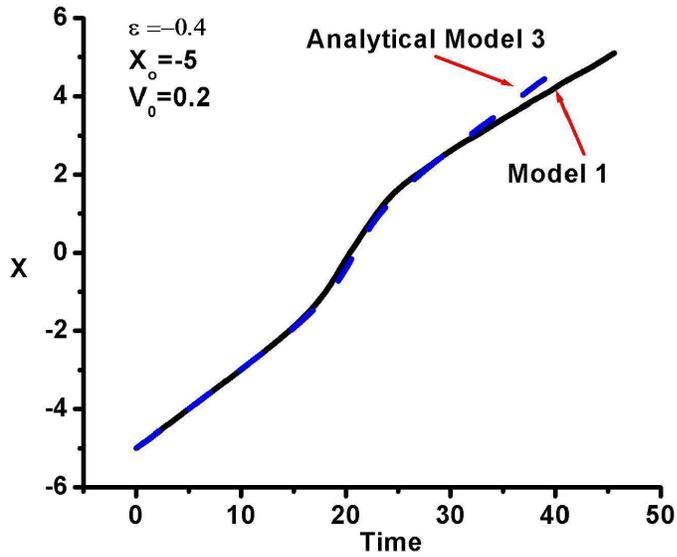}
  \end{center}
\caption {Trajectory of a soliton during the interaction with a potential well. Solid line shows the simulation results using model 1 and dashed lines presents the results of the model 3. }
  \label{fig:fig8}
\end{figure}

A soliton with a low initial velocity might get trapped by the potential and oscillate in the well. The Period of the oscillation can be calculated with model 3 from (24) with a negative potential strength $\left(\epsilon<0\right)$.

Figure 9 demonstrates an oscillating situation. This figure presents the trajectory of a soliton in a potential well with $\sigma=-0.4$ for the model 1 ($\epsilon_{effective}=-0.29$ for model 3). The soliton is located at the initial position $X_{0}=-3$ and starts moving with an initial velocity $\dot{X_{0}}=0.01$. The period of the oscillation simulated by model 1 is , $T\approx398$ while the period calculated by (27) is about 372.

\begin{figure}[htbp]
  \begin{center}
    \leavevmode
  \epsfxsize=10cm
  \epsfbox{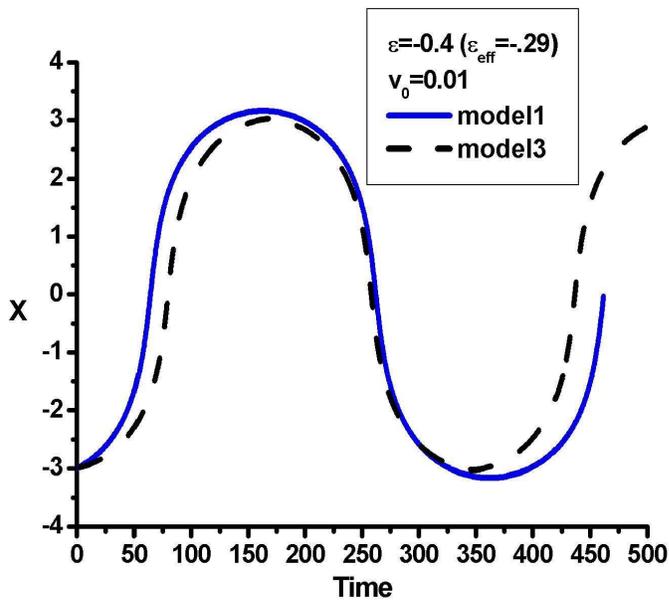}
  \end{center}
\caption {Oscillation of a soliton in a potential well. Solid line presents the results simulation using model 1and dashed line shows the trajectory of the soliton calculated with model 3. }
  \label{fig:fig9}
\end{figure}

Due to using adiabatic approximation in model 3, the results show noticeable differences among the models when the velocity has rapid or big changes. In a situation where a soliton moves from an initial position very far from the potential with a very low velocity the models fail to match by using the fitting equations (21) and (22). This means that a better approximation is needed, but the analytic model is acceptable.

\begin{figure}[htbp]
  \begin{center}
    \leavevmode
  \epsfxsize=10cm  \epsfbox{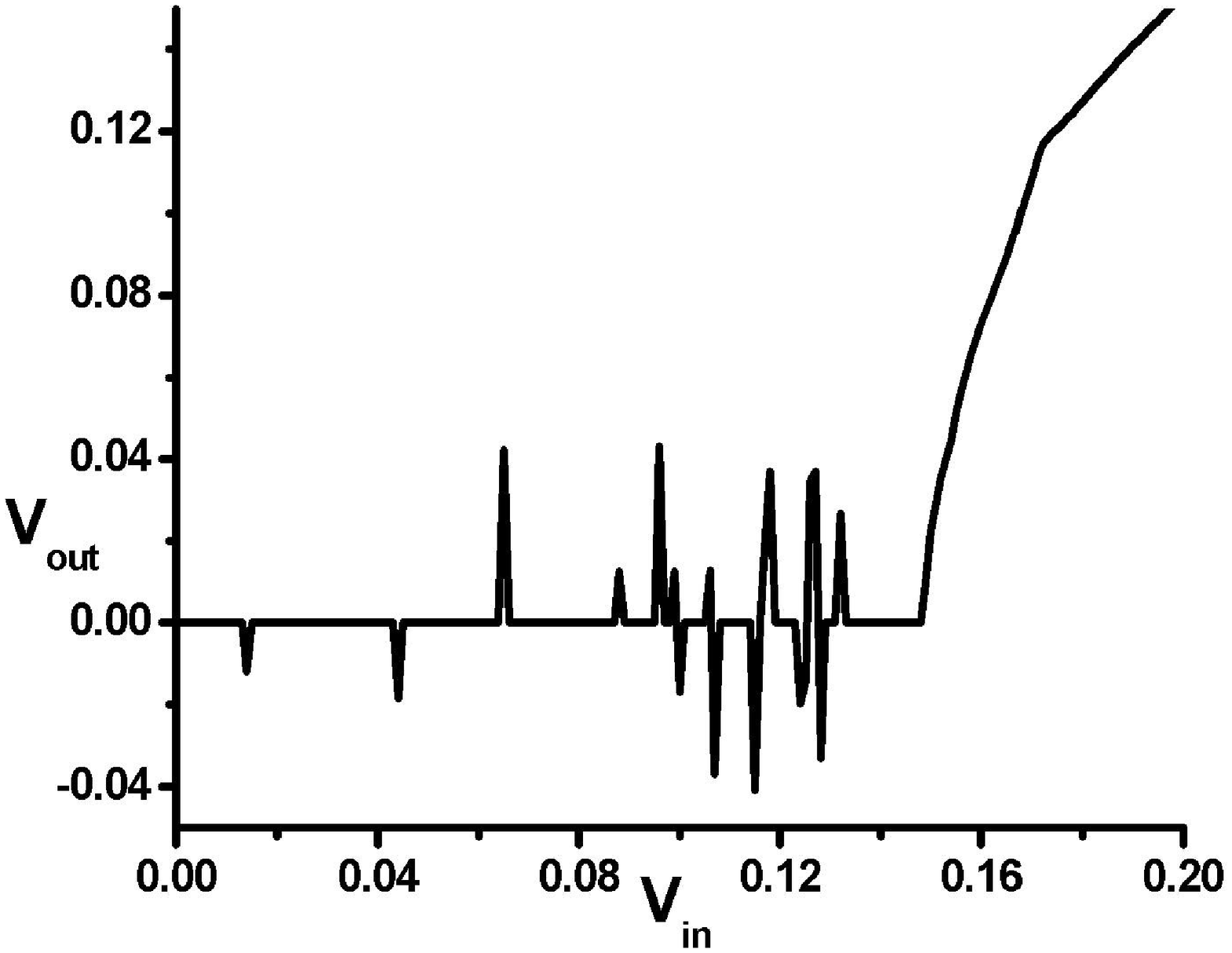}
  \end{center}
\caption {Final soliton velocity as a function of its initial velocity for $\epsilon=-0.5$. Zero final velocity means that the soliton is captured by the potential well.}
  \label{fig:fig10}
\end{figure}

An attractive situation in the soliton-well interaction is the fine structure of the islands of trapping. In model 1, the final situation of a soliton with an initial velocity lower than the critical velocity is very sensitive to its initial conditions and the strength of the well. In most of the cases when the incoming velocity of the soliton is smaller than critical velocity the soliton cannot escape from the potential. Particularly, after the first interaction soliton will stop and then it will return to interact with the potential again. For most of the initial conditions, the soliton will lose its energy again during the following interactions and finally becomes trapped by the potential. Equation (18) of model 3 clearly shows that the effective force is a function of the soliton velocity, so it may be dissipative. However, for some specific initia1 velocities, the soliton may escape to ($\pm\infty$) after some interactions. Finding an analytical description for the windows of trapping in the model 3 is hard and we couldn't find a clear analytical formulation for this phenomenon using model 3. But this situation can be studied with solving equation (18) numerically. For investigating of this phenomenon in the model 3, a plot of "initial velocity respect to final velocity" of the soliton is needed. Figure 10 shows the outgoing velocity of the soliton as a function of its incoming velocity. The initial position of $X_{0}=-2$ was used in simulations. The outgoing velocity has been calculated when the soliton reaches $X=\pm10$. Zero final velocity means that the soliton is captured by the potential. The differential equation (18) has been integrated numerically by using "quality-controlled" step size in Runge-Kutta method with maximum error less than 0.001. Simulations at this precision show reflection and also transmission from the potential well at some initial velocities. We have to check the validity of the results by examining the windows of escaping of figure 10. For this purpose we need the trajectory of the soliton with an initial velocity in the region of escaping, but with a higher precision. Some simulations with better precisions have been performed using Maple which contains some advanced algorithm with higher precision. The results show that for most of the escaping windows the soliton trapped in the well if we simulate the situation with better approximation. This means that this situation is very sensitive to the precision of numerical calculations. However there are some values of initial velocities in which the soliton can transmit over the well or may reflect back.This phenomenon needs a deeper investigation.

\section{Conclusion and Remarks}

In this article, an analytical model for the soliton-potential interaction has been presented. It is shown that the model has a very close relation with other models in the way that it is possible to fit this model over the other models. The model gives the critical velocity in the soliton-potential interaction as a function of initial conditions of the soliton and the characters of the potential. The model predicts specific relations between some functions of initial conditions and other functions of final state of the soliton during the interaction. Also the model presents a good approximation for the trajectory of the soliton during the interaction. The oscillation period of the soliton in the well can also be calculated by the analytical model. Simulations using other models are in agreement with the present analytic model. But this model does not predict the narrow windows of soliton reflection from the potential well.

The model can be used for prediction the results of other potentials beside the sine-Gordon model.

{\bf Acknowledgments}:

Author is grateful to A.R. Mokhtari and A.R. Etezadi for discussions and helps.


\end{document}